# Regional differences in subduction ground motions

**C. Beauval[1], F. Cotton[1], N. Abrahamson[2], N. Theodulidis[3], E. Delavaud[4], L. Rodriguez[1], F. Scherbaum[5], A. Haendel[5]**
[1] ISTerre, Grenoble, France
[2] Pacific Gas and Electric, San Francisco, US
[3] ITSAK, Thessaloniki, Greece
[4] ETH, Zurich, Switzerland
[5] University of Potsdam, Germany

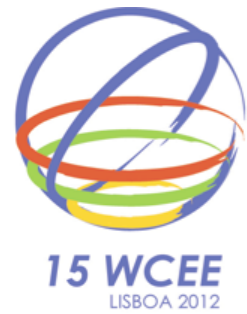

**SUMMARY:**
A few ground-motion prediction models have been published in the last years, for predicting ground motions produced by interface and intraslab earthquakes. When one must carry out a probabilistic seismic hazard analysis in a region including a subduction zone, GMPEs must be selected to feed a logic tree. In the present study, the aim is to identify which models provide the best fit to the dataset M6+, global or local models. The subduction regions considered are Japan, Taiwan, Central and South America, and Greece. Most of the data comes from the database built to develop the new BCHydro subduction global GMPE (Abrahamson et al., submitted). We show that this model is among best-fitting models in all cases, followed closely by Zhao et al. (2006), whereas the local Lin and Lee (2008) is well predicting the data in Taiwan and also in Greece. The Scherbaum et al. (2009) LLH method prove to be efficient in providing one number quantifying the overall fit, but additional analysis on the between-event and within-event variabilities are mandatory, to control if median prediction per event and/or variability within an event is within the scatter predicted by the model.

## 1. INTRODUCTION

To carry out a probabilistic seismic hazard calculation for a region, an empirical ground-motion prediction equation (GMPE) is needed to predict ground motions generated by earthquakes. The GMPE selected might have been developed from local data, but more often due to the lack of data a model including data from different regions is used. Few regions in the world have a strong-motion database complete enough to be used alone for deriving a prediction equation. The present study focuses on empirical GMPEs developed for predicting the ground motions produced by subduction interface and inslab events. Inslab events produce on average larger near-source amplitudes than interface events, but these amplitudes attenuate much faster, causing damaging ground motions over a much smaller area. All the data considered here (except for Greece) is extracted from the generating dataset of the global model developed by Abrahamson et al. (submitted), gathering data from different subduction zones (Japan, Taiwan, Central and South America, ..).
We test candidate GMPEs developed for subduction zones at a global or local scale against observations. The "global" GMPEs have been developed based on worldwide data assuming that there is no major regional difference in the attenuation of ground motions between subduction zones (Youngs et al. 1997, Atkinson and Boore 2003, Abrahamson et al. submitted). Other models are regional, established using mainly local data, and aimed at predicting accelerations for a specific subduction region (e.g. Zhao et al. 2006, Lin and Lee 2008, Arroyo et al. 2010).
The aim of the present study is to understand the performance of global and local models with respect to the data of the different seismic zones. The databases used for the regression of the GMPEs may differ or partially overlap the dataset considered here, and thus we do not expect a perfect adequacy between models and data. To quantify the fit, the LLH method proposed by Scherbaum et al. (2009) is used. The LLH evaluates the overall performance of the model with respect to the data. Complementary to the LLH values, event terms and within-event dispersions are calculated for the earthquakes considered. We will focus on the earthquakes contributing the most to the seismic hazard in these regions, i.e. moment magnitudes higher or equal to 6. Only recordings for which spectral accelerations can be calculated up to at least 25 Hz are selected.

## 2. METHOD

Scherbaum et al. (2009) provides a ranking criterion based on information theory (see the original paper for a detailed description of the theory behind). This technique is based on the probability for an observed ground motion to be realized under the hypothesis that a model is true. It provides one value, the negative average log-likelihood LLH (Delavaud et al. 2012), that reflects the fit between data and model:

$$\text{LLH} := -\frac{1}{N}\sum_{i=1}^{N} \log_2(g(x_i)).$$

with N the number of observations $x_i$, and g the probability density function predicted by the GMPE (normal distribution, with standard deviation the total sigma of the model). The ranking of models according to their fit to the data is then straightforward. A small LLH indicates that the candidate model is close to the model that has generated the data, while a large LLH corresponds to a model that is less likely of having generated the data. Taking advantage of synthetic data, the LLH value can be easily interpreted in terms of residual distributions (Fig. 1, see Beauval et al. in press). Synthetic datasets are generated from an original Gaussian distribution, and distributions with modified characteristics (in terms of mean and sigma) are tested against these synthetic datasets. The final LLH characterizing the fit between the data and the model is simply the mean of all individual LLH. If the tested distribution is equal to the original distribution, the LLH value is close to 1.4-1.5. If the tested Gaussian is different from the original distribution, the LLH increases.

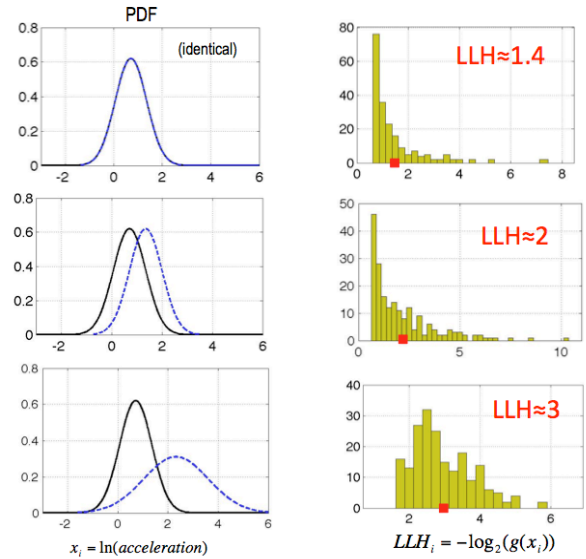

**Figure 1.** LLH calculation: synthetic ground motion datasets generated from an original Gaussian distribution (black), are tested against distributions with modified characteristics (dashed blue curve). The final LLH is the mean of the 200 LLH values calculate from the 200 synthetic accelerations.

Besides LLH values, we calculate event terms and within-event variabilities, and compare them to the between-event and within-event sigmas of the GMPEs. The between-event dispersion integrates the variability on the characteristics of the source, whereas the within-event dispersion integrates roughly the variability in the wave propagating path and in the site condition. The total residual $R_i$ between an observed acceleration and the acceleration predicted by a particular GMPE is obtained as follows:

$R_i = \ln(A_i)_{OBS} - \ln(A_i)_{GMPE}$

where '$_{OBS}$' refers to the observed acceleration from recording i, and '$_{GMPE}$' refers to the median acceleration value, for location i predicted by the GMPE, considering the earthquake magnitude, appropriate source distance and site condition (5% damped spectral acceleration). The event term of an event is simply the mean value of the residuals, therefore calculated over all locations available for the event. The event terms can then be compared to the between-event sigma ($\tau$) of the GMPE, which has been evaluated empirically from recorded earthquakes during the development of the GMPE. If the event terms is small and fall within one standard deviation, this suggests that the model predicts the records rather well. The within-event variabilities are also calculated, they correspond to the standard deviation of the residuals of each earthquake. They can be compared to the within-event sigma of the model ($\phi$).

## 3. COMPARISON OF PREDICTIONS AND OBSERVATIONS

### 3.1 Ground-motions from interface and inslab events in Japan

The earthquakes considered here are included in the generating dataset of Zhao et al. 2006 model, which is constituted of all available records from earthquakes with moment magnitudes higher or equal to 5.0 since 1968 in Japan.

INTERFACE: We consider 7 interface earthquakes that occurred in the two subduction zones in Japan (Table 1 and Fig. 2, in total 800 recordings with source distances up to 300km). The LLH values, reflecting the fit between the predictions of the model and the observations, varies with the GMPE considered, but are quite stable according to frequency for a given GMPE. The GMPEs predicting accelerations which are the most in accordance with the recordings are 1) the global model BCHydro (Abrahamson et al. submitted), 2) Zhao et al. (2006) established from a Japanese database (a few overseas events at close distance), 3) Kanno et al. (2006) established from Japanese data (behavior for f<2.5Hz needs to be understood). The Lin and Lee model (2008), established for Taiwan, from both local data (up to $M_w$=6) and foreign data (M>6), also provides reasonable LLH values, although slightly higher. Youngs et al. (1997) and Atkinson and Boore (2003) models are yielding higher LLH (2.2-3.2).

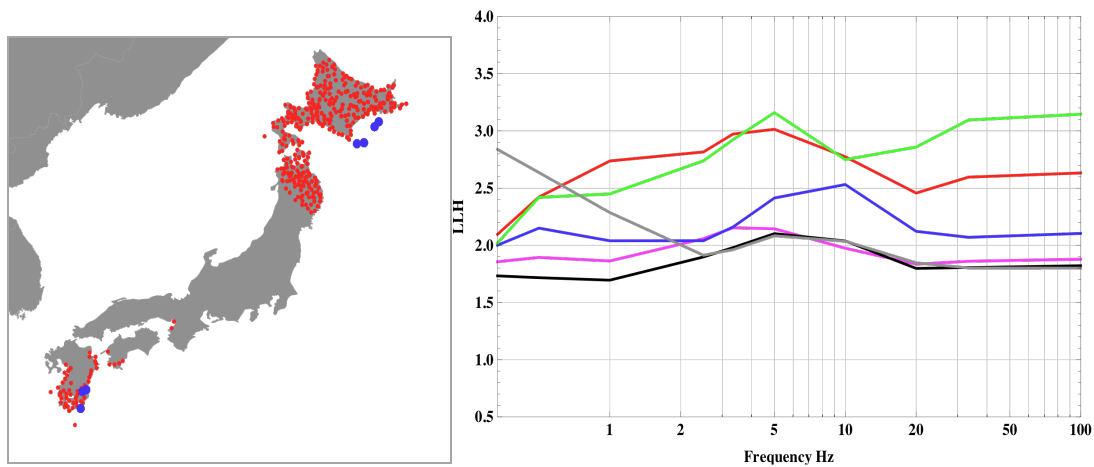

**Figure 2.** Japan, interface events considered: map of stations (red points) and epicenters (blue points), and LLH values versus frequency, calculated for the GMPEs: BCHydro, Zhao et al., Lin and Lee, Kanno et al., Youngs et al., Atkinson and Boore.

**Table 1.** Japan: interface and inslab events used in the testing (parameters from Abrahamson et al. (submitted))

| Date | Magnitude Mw | Depth (km) | # of recording sites | Source type |
|---|---|---|---|---|
| 1996/10/19 | 6.7 | 18 | 47 | Interface |
| 1996/12/03 | 6.7 | 29 | 33 | Interface |
| 2000/06/25 | 6. | 10 | 25 | Interface |
| 2003/09/26 | 8.3 | 23 | 319 | Interface |
| 2003/09/26 | 7.4 | 50 | 222 | Interface |
| 2003/09/25 | 6.5 | 41 | 86 | Interface |
| 2003/10/08 | 6.7 | 41 | 68 | Interface |
| 1996/09/11 | 6.1 | 58 | 40 | Intraslab |
| 1999/01/24 | 6.4 | 45 | 33 | Intraslab |
| 2000/01/28 | 6.8 | 53 | 36 | Intraslab |
| 2000/06/03 | 6.1 | 50 | 76 | Intraslab |
| 2001/03/24 | 6.8 | 53 | 67 | Intraslab |
| 2001/12/02 | 6.5 | 123 | 97 | Intraslab |
| 2003/05/26 | 7. | 77 | 122 | Intraslab |

The event term and the within-event dispersion for each earthquake are calculated relatively to the Abrahamson et al. (2010) GMPE, and compare to the corresponding sigmas of the model (Fig. 3, see e.g. Stewart et al. 2012). It is interesting to observe that the event terms of the two earthquakes contributing the most to the LLH value (Table 1, 222 and 319 recordings each) are yielding opposite trends. For one earthquake, predictions are under-predicting the observations and event terms range from 0.3 to 0.6 (natural log units), while event terms range from -0.7 to -0.4 for the other event. Considering all earthquakes, event terms are roughly within the scatter of event terms from the earthquakes used in the development of the BCHydro equation. The within-event variabilities are superimposed to the within-event sigma of the model (Fig. 3 right). Depending on the event, the dispersion is equal, slightly higher or lower than the within-event sigma.

The same analysis is performed for understanding the results obtained for the Zhao et al. (2006) model (Fig. 4), which is yielding LLH values slightly higher than the BCHydro model. The event terms for the 6 earthquakes range between $+1\tau$ and $-2\tau$, indicating that the ground motion depends greatly on the specific earthquake, and that accelerations are on average lower than predicted by the model. On the other side, the model is rather well predicting the within-event variability (Fig. 6, right).

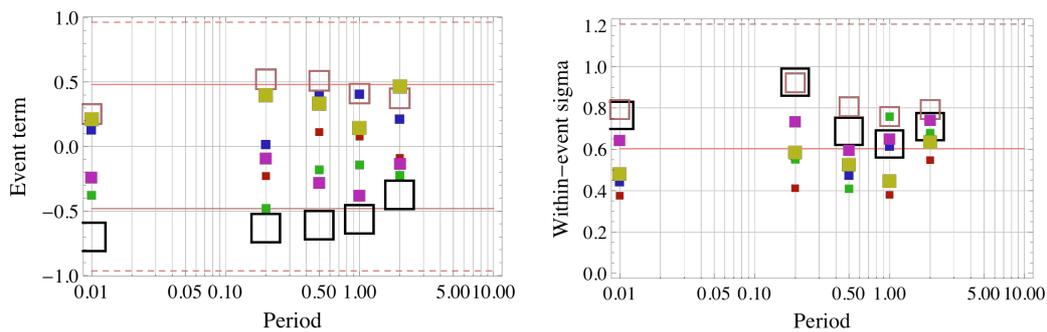

**Figure 3.** Data: Japan interface events (Table 1), GMPE: BChydro (Abrahamson et al. submitted). Left: event terms for each event, calculated for 5 periods: 0.01 (≈PGA), 0.2, 0.5, 1, 2 seconds. The size of the symbol is proportional to the number of recordings per event. Red solid lines: $\pm\tau$ (sigma between-event of the GMPE); red dashed lines: $\pm 2\tau$. Right: dispersion within each event (sigma of the residuals). Red line: $\phi$ (sigma within-event of the GMPE); red dashed line: $+2\phi$.

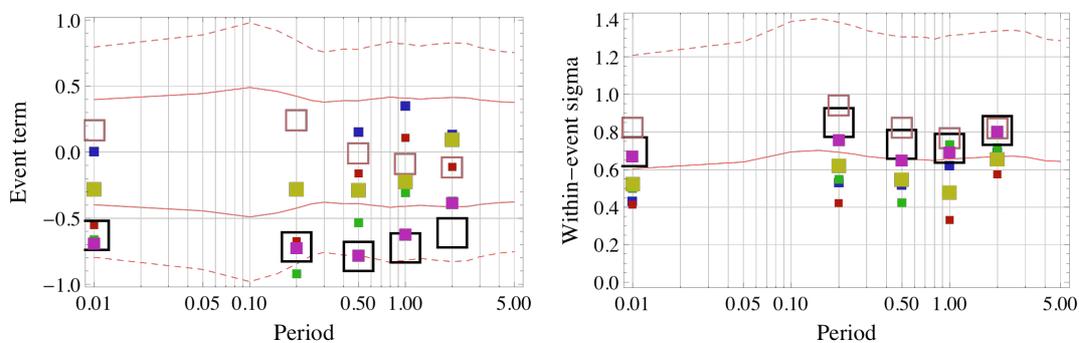

**Figure 4.** Data: Japan interface, GMPE: Zhao et al. (2006). See legend of Fig. 3. The size of the symbol is proportional to the number of recordings per event.

INTRASLAB: We consider 7 intraslab earthquakes that occurred in the two subduction zones in Japan

(Table 1 and Fig. 5, in total 844 recordings). The obtained LLH values rank first the local model Zhao et al. (2006), then comes Kanno et al. (2006), as well as the global models Abrahamson et al. (2010) and Atkinson and Boore (2003). Note that, unlike Atkinson and Boore (2003) who discarded distances beyond 200km for the Japanese recordings, here sources distances up to 300 km are taken into account. Youngs et al. (1997) model is yielding higher LLH values (2-2.5), whereas Lin and Lee (2008) and McVerry et al. (2006) are predicting accelerations which do not fit the dataset.

The event terms relative to the best fitting equation, Zhao et al., are roughly included in the typical range of past Japanese earthquakes (varying between -1τ and +2τ, Fig. not shown), whereas the within-event standard deviation of the 7 earthquakes are very close to the within-event sigma of the model.

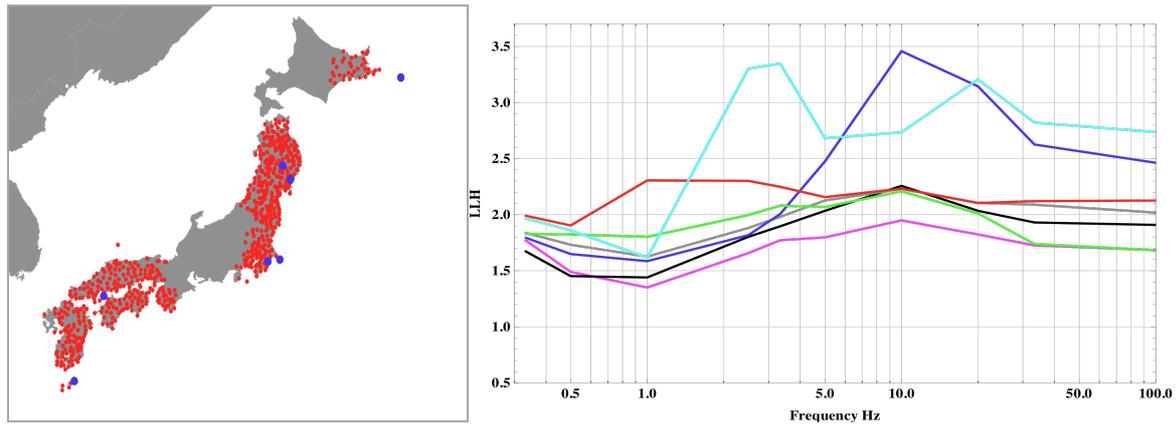

**Figure 5.** Japan, intraslab events considered: map of stations and epicenters, and LLH values versus frequency, calculated for the GMPEs: BCHydro, Zhao et al., Lin and Lee, Kanno et al., Youngs et al., Atkinson and Boore, McVerry et al.

### 3.2. Ground-motions from interface and inslab events in Taiwan

INTERFACE: Three of the six interface earthquakes considered are included in the dataset used in the regression of the Lin and Lee (2008) GMPE for northeastern Taiwan (events with M≥7 are all foreign events). The six interface earthquakes are located on the same subduction zone, at the boundary between the Philippine Sea plate and the Eurasian plate.

Results are displayed in Fig. 6. Above 2 Hz, the models yielding the best fit to the data are the BCHydro and Lin and Lee (2008) models. The models Zhao et al. (2006), Atkinson and Boore (2003) and Kanno et al. (2006) yield slightly higher LLH. Youngs et al. and McVerry et al. are predicting accelerations which differ significantly from the observations (2<LLH<3.3). Below 2 Hz, none of the model is providing a good fit. This poor fit for long periods will need to be understood.

Calculating the event terms relative to the Lin and Lee model show the same trend for all earthquakes: the median of the residuals for one earthquake is in the scatter predicted by the model for periods lower than 0.5s (event terms between -σ and 0, Fig. not shown). Note that in Lin and Lee (2008), only the total sigma (σ) is provided. For higher periods, the event terms takes increasingly positive values. As for the dispersion quantified within each earthquake, it is close to the variability predicted by the model (and thus close to the variability of the generating dataset of Lin and Lee (2008) model).

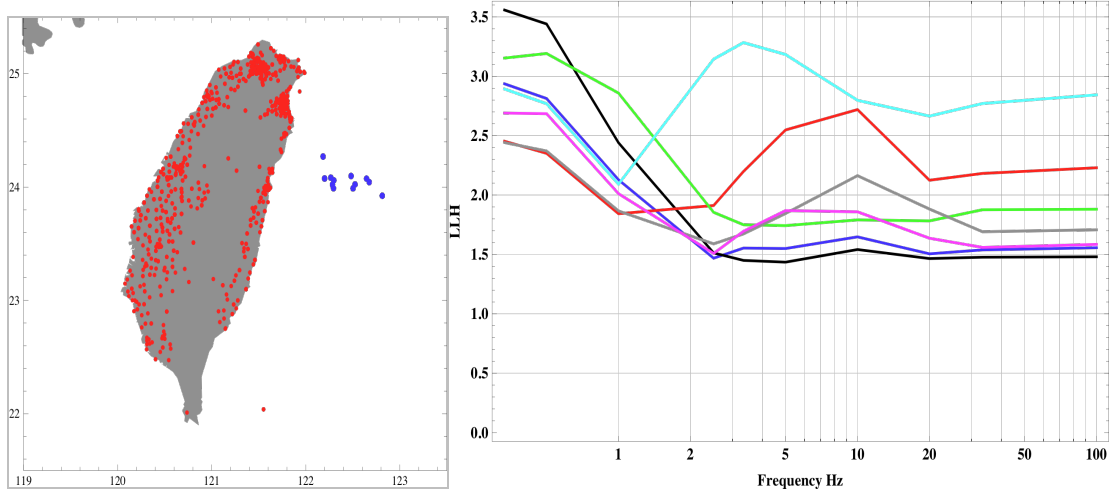

**Figure 6.** Taiwan, interface events: map of stations and epicenters, and LLH values versus frequency, calculated for the models: BCHydro, Zhao et al., Lin and Lee, Kanno et al., Youngs et al., Atkinson and Boore, McVerry et al.

**Table 2.** Taiwan: interface and inslab events used in the testing (parameters from Abrahamson et al. (submitted))

| Date | Magnitude Mw | Depth (km) | # of recording sites | Source type |
|---|---|---|---|---|
| 1992/09/28 | 6.4 | 38 | 10 | Interface |
| 1994/05/23 | 6.2 | 32 | 23 | Interface |
| 1996/03/05 | 6.3 | 44 | 207 | Interface |
| 2002/03/31 | 7.1 | 32 | 381 | Interface |
| 2001/12/18 | 6.8 | 28 | 308 | Interface |
| 2004/11/08 | 6.3 | 30 | 271 | Interface |
| 1995/06/25 | 6. | 57 | 275 | Intraslab |
| 2004/10/15 | 6.6 | 112 | 434 | Intraslab |
| 2005/10/15 | 6.4 | 200 | 176 | Intraslab |
| 2006/12/16 | 7. | 44 | 309 | Intraslab |
| 2006/12/26 | 6.7 | 50 | 475 | Intraslab |

INTRASLAB: One of the five intraslab earthquakes considered are included in the generating dataset of the Lin and Lee (2008) GMPE developed for northeastern Taiwan (event in 1995). In the generating dataset, events with M>6 are all foreign events.

For most models (Fig. 7), the resulting LLH values are stable according to the frequency. The best-fitting models are the Zhao et al. (2006) Japanese model, the Lin and Lee (2008) model developed for Taiwan, and the BCHydro model (1.5<LLH< 2.5). McVerry et al. (2006), Youngs et al. (1997) and Atkinson and Boore (2003) models yield comparable LLH values above 5 Hz, but higher values for lower frequencies. The event terms are estimated relatively to the Zhao et al. (2006) model, and compared to its between-event sigma (Fig. 8, left). Event terms vary according to the earthquake, negative for some, positive for others, but contained in the scatter predicted by the model (within -τ and +2τ). The variability of the ground motion within each earthquake is very close to the within-event sigma of the model (Fig. 8, right).

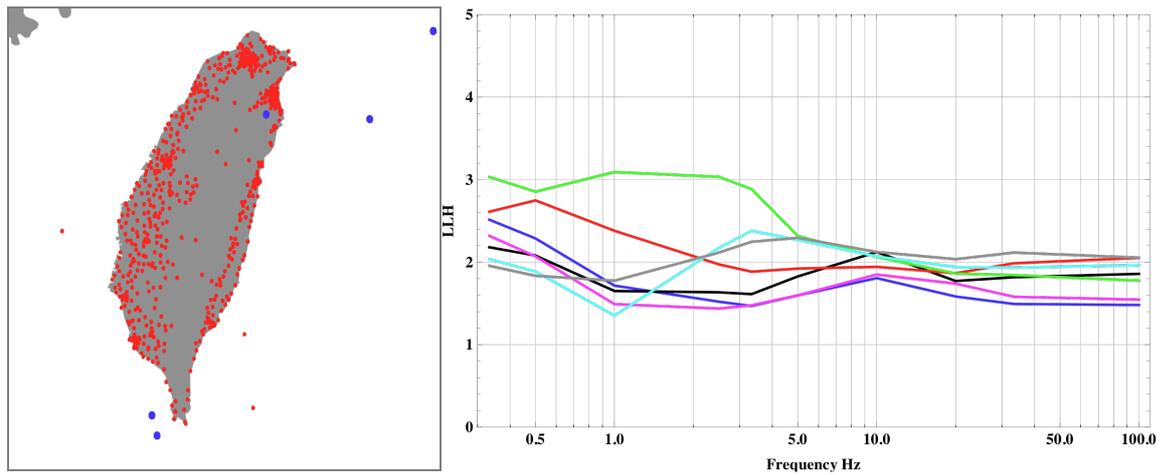

**Figure 7.** Taiwan, intraslab events considered: map of stations and epicenters, and LLH versus frequency, for the GMPEs: BCHydro, Zhao et al., Lin and Lee, Kanno et al., Youngs et al., Atkinson and Boore, McVerry et al.

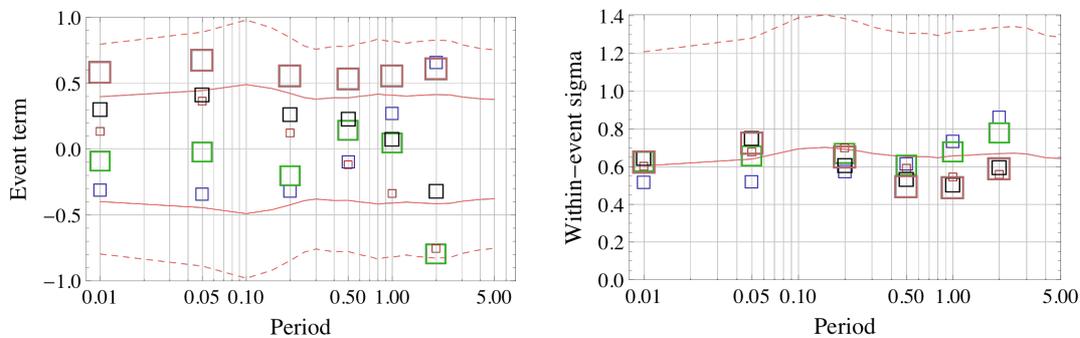

**Figure 8.** Data: Taiwan intraslab events, GMPE: Zhao et al. (2006). See legend of Fig. 3. The size of the symbol is proportional to the number of recordings per event.

### 3.3 Ground-motions from interface events in Central and South America

From the global database, 15 interface events have been extracted, with moment magnitude higher or equal to 6, and spectral accelerations extending at least up to 25 Hz, providing in total 113 recordings (rupture distance up to 300km). We consider as a whole central and South America, although we might be mixing different interface geometries.
As shown in Fig. 9, the model ranked first and yielding lowest LLH values is the Arroyo et al. (2010) GMPE, designed from and for Mexican interface events (on rock sites). This result must be taken with caution, as the number of recordings is divided by two because only records at rock are considered. Models by Atkinson and Boore (2003), Zhao et al. (2006), BCHydro, and Youngs et al. (1997) appear to well predict the data. They yield close LLH values (in-between 1.5 and 1.8) stable with frequency. The results are quite coherent with the study of Arango et al. (2012), which identify the same 4 best-fitting models from a larger dataset of interface events in the Peru-Chile region. In the same study, they find that Lin and Lee (2008) model is well predicting data of interface events in Central America, which is not the case on our dataset.

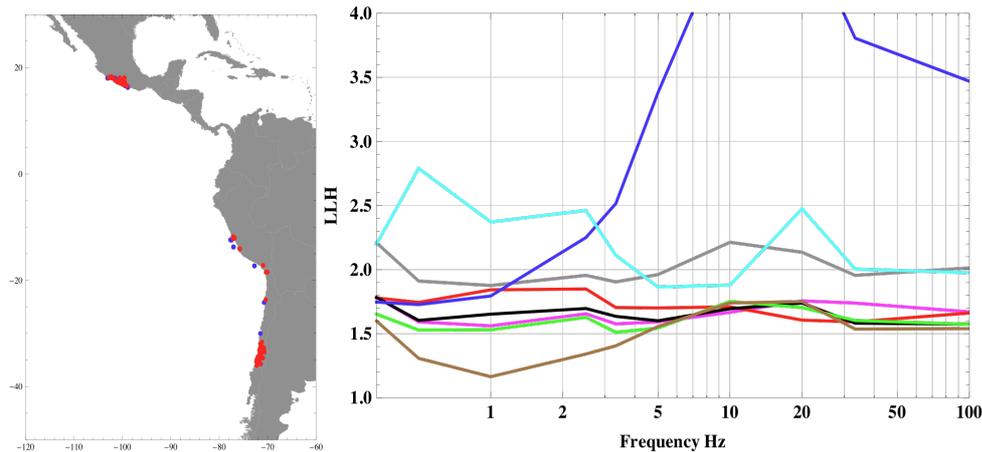

**Figure 9.** Central and South America, interface events: map of stations and epicenters, and LLH versus frequency calculated for the GMPEs: BCHydro, Zhao et al., Lin and Lee, Kanno et al., Youngs et al., Atkinson and Boore, McVerry et al, Arroyo et al. (only the 54 recordings on rock sites have been used in this case).

### 3.4 Ground-motions from inslab events in Greece

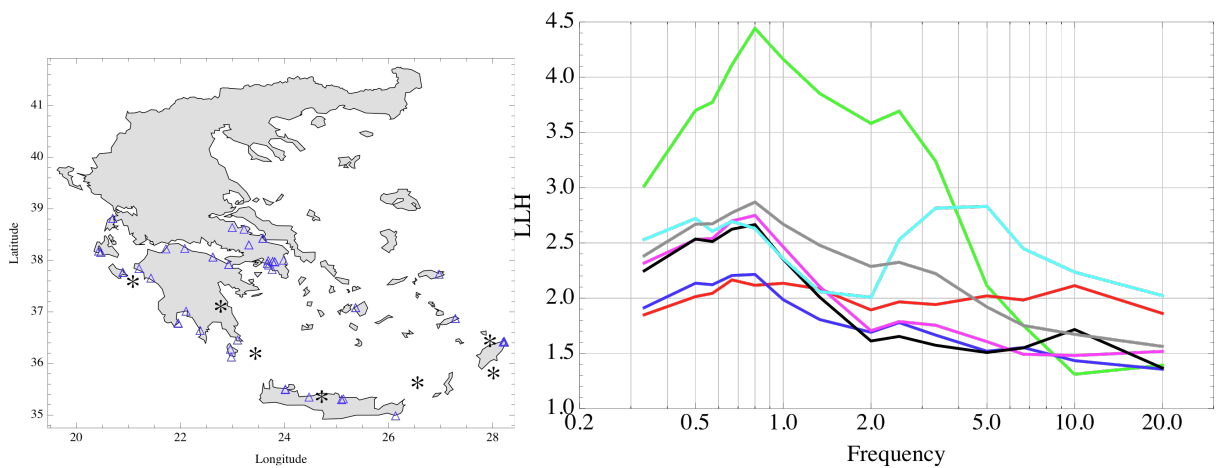

**Figure 10.** Greece, intraslab events: map of stations and epicenters, and LLH values versus frequency, calculated for the GMPEs: BCHydro, Zhao et al., Lin and Lee, Kanno et al., Youngs et al., Atkinson and Boore, McVerry et al.

The Greek subduction zone is the most active subduction zone within Europe. Intermediate-depth earthquakes along the Hellenic arc occur along a well-defined Wadati-Benioff zone, which have more or less amphitheatrical shape, at depths ranging from about 60 to 170km. The Wadati-Benioff zone is due to the subduction of the Eastern Mediterranean lithosphere under the Aegean microplate. Intermediate-depth events are strike-slip events with a significant thrust component. The acceleration dataset was compiled using data recorded by the acceleration-sensor networks operated by the Institute of Engineering Seismology and Earthquake Engineering (ITSAK). Source-site distances range from 70 to 300km.

Currently, there is no local ground-motion prediction model fulfilling all the requirements for being used in a PSHA study. In the SHARE project (Delavaud et al. 2012), GMPEs needed to be selected to predict the ground motions produced by interface and intraslab earthquakes. The choice of the GMPEs was guided both by expert opinion and using the results of the testing on the data available. The data available is restricted (Table 3, 7 events and 68 recordings), but still it is worth evaluating the fit with respect to the local and global models.

Three models are providing a relatively good fit to the data for frequencies higher than 2Hz:

BCHydro, Zhao et al. (2006) and Lin and Lee (2008). For the same frequency range, Kanno et al. (2006), Youngs et al. (1997), and McVerry et al. (2006) are providing higher LLH (between 1.6 and 2.8), while Atkinson and Boore (2003) yield large LLH (up to 3.6). Below 2 Hz, LLH values of all models increase. Lin and Lee and Youngs et al. models are best predicting the observations, with LLH around 2.0.

**Table 3.** Greece: inslab events used in the testing

| Date | Magnitude Mw | Depth (km) | # of recording sites |
|---|---|---|---|
| 1994/05/23 | 6.1 | 90 | 4 |
| 1996/04/26 | 5.8 | 64 | 2 |
| 1999/06/11 | 5.2 | 91 | 6 |
| 2006/01/08 | 6.7 | 66 | 19 |
| 2008/01/06 | 6.2 | 80 | 22 |
| 2008/07/15 | 6.4 | 40 | 7 |
| 2011/04/01 | 6.1 | 60 | 8 |

## 4. DISCUSSION AND CONCLUSIONS

The aim of the study was to test systematically global and local models against the same regional datasets (M≥6, most relevant to hazard calculation), and obtain a hierarchy of their fit to the data. Among the global models, the BCHydro equation (Abrahamson et al., submitted) appears to be the model which is providing the best fit to interface and intraslab datasets, in the different regions tested (Japan, Taiwan, Greece, Central and South America, Greece). The good performance of this model, with respect to the local models, does not highlight strong regional differences, suggesting that it can be applied safely for predicting ground motions in all subduction regions. Among the local models, the Zhao et al. model (2006, Japan) shows to predict quite well the data, in all regions studied. The Lin and Lee (2008) model (Taiwan) is predicting well the ground motions produced by earthquakes in Taiwan and in Greece, but not elsewhere. Atkinson and Boore (2003) global base model is used here (the correction factors for Japan are not accounted for). It provides rather good fit to the data, except in the case of interface events in Japan and inslab events in the Greek subduction zone.

The between-event and within-event variabilities, playing a role in the LLH estimate, must be partitioned. Analysing the event terms and within-event variabilities relative to a GMPE indicates if the event terms for specific regions are consistent with event-to-event scatter predicted by the models (observed from past earthquakes). The within-event variabilities of the earthquakes can be compared to the within-event sigma of the models. We have analysed the residuals with respect to all models, but only a few examples were displayed in the paper. They show that a small LLH usually correspond to within-event variabilities compatible with the within-event sigma of the model, and to event terms which are mostly contained in the scatter predicted by the model (negative and positive, and resulting in an average close to 0). More work is required to have a more comprehensive picture of the adequacy between models and data, which has been evaluated here in a broad sense, in particular in terms of attenuation of amplitudes with distance. More work is also needed on analysing the regional variability of the ground motions on a finer scale and its dependence with frequency.

**AKCNOWLEDGEMENT**
This work has been done mainly with the support of the SHARE European program FP7 (contract number 226967, http://www.share-eu.org/).